\documentclass[twocolumn,prb]{revtex4}

\usepackage{graphics}

\usepackage{epsfig} 

\usepackage{dcolumn}

\usepackage{amsmath}

\begin{document}

\title{ Conductance of ion channels and nanopores with charged walls: a toy model }

\author{ J.~Zhang, A.~Kamenev, B.~I.~Shklovskii}

\address{Department of Physics, University of Minnesota,
Minneapolis, MN 55455, USA}

\date{\today}

\begin{abstract}

We consider ion transport through protein ion channels in lipid
membranes and  water--filled nanopores in silicon films. It is
known that, due to the large ratio of dielectric constants of
water and the surrounding material, an ion placed inside the
channel faces a large electrostatic self--energy barrier. The
barrier leads to an exponentially large resistance of the channel.
We study reduction of the electrostatic barrier by immobile
charges located on the internal walls of the channel. We show that
the barrier practically vanishes already at relatively small
concentration of wall charges.

\end{abstract}

\maketitle

\maketitle

\begin{figure}[ht]

\begin{center}
\includegraphics[height=0.17\textheight]{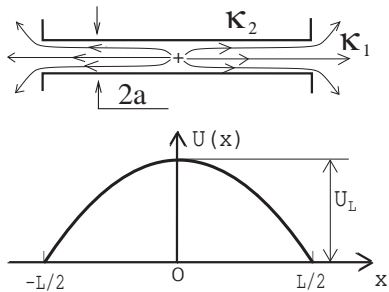}
\end{center}

\caption{ Electric field of a cation in a short cylindrical
channel with the large dielectric constant $\kappa_1\gg \kappa_2$.
$L$ is the channel length, $a$ is its radius. The self--energy
barrier is shown as a function of the coordinate $x$. }
\label{fig1}
\end{figure}

Protein ion channels functioning in biological lipid membranes is
a major frontier of biophysics ~\cite{Hille,Doyle}. An ion channel
can be inserted in an artificial membrane in vitro and studied by
physical methods. Similar artificial devices -- water--filled
nanopores, are studied in silicon, silicon oxide films and polymer
membranes~\cite{Li}. In both cases, one can study a single water
filled channel connecting two reservoirs with salty water
(Fig.~\ref{fig1}). A static voltage applied between these
reservoirs drops almost entirely in the channel due to the high
conductivity of the bulk solution. The voltage drives salt cations
and anions through the channel. One can measure the ohmic
resistance of the channel.

This resistance may be exponentially large due to the fact that
the dielectric constant of water $\kappa_1 \simeq 80$ greatly
exceeds that of the surrounding media $\kappa_2$ ($\kappa_2\simeq
2$ for lipids and $\kappa_2\simeq 4$ for silicon oxide). Indeed,
in this case the electric field of an ion traversing the channel
is forced to stay inside the channel (Fig.~\ref{fig1}). This
creates a barrier $U(x)$, where $x\in[-L/2,L/2]$ is the ion
coordinate inside the channel. The barrier is the difference
between the self--energy of the ion at the point $x$ inside the
channel and the self--energy in the bulk~\cite{Parsegian}. It is
the maximum of the barrier, $U(0)=U_L$, that determines the
resistance of the channel.
If a channel is very long the electric lines leak through the
protein walls and lipids so that the barrier saturates as a
function of $L$. Roughly speaking this
happens~\cite{Parsegian,Fink1,Kamenev} at $L \sim
a(\kappa_1/\kappa_2)^{1/2}$, where $a$ is the channel radius. In
this paper we assume that the channel is shorter, so that we can
neglect the field leakage. In this case, calculation of the
barrier height is very simple~\cite{Fink1,Kamenev}. The electric
field at a distance $x > a$ from a cation located in the middle of
the short channel is uniform and and according to the Gauss
theorem is $E_0 = 2e/\kappa_1 a^{2}$. The energy of such field in
the volume of the channel is
\begin{equation}
U_L(0) = {\kappa_1 E_{0}^{2}\pi a^{2}L\over 8\pi} = {e^{2}L\over
2\kappa_1 a^{\, 2}}={eE_{0}L\over 4}\, , \label{short}
\end{equation}
where the zero argument is added to indicate that there are no
other charges in the channel. $U_{L}(0)$ is proportional to  $L$
and (for a narrow channel) can be much larger than $k_BT$, making
the channel resistance exponentially large.

At large concentration of salt in surrounding water the
electrostatic barrier is reduced by
screening~\cite{Jordan1,Kamenev}. In biological channels the
nature uses more effective approach. Channels designed for the
transport of cations (K, Na, Ca) have negative charges on internal
walls. For example, the potassium channel has 8 amino-acids with
negatively charged radicals build into the wall of the
protein~\cite{Doyle,Hille}. Walls of artificial nanopores,
generally speaking, are charged as well and one can control these
charges by a chemical treatment and/or  tuning   pH of the
solution. The goal of this paper is to study the effect of
immobile wall charges on the electrostatic barrier and the channel
resistance. For certainty we assume that wall charges are negative
and equally spaced  along the channel  with the  linear density
$n_w$.

\begin{figure}[ht]
\begin{center}
\includegraphics[height=0.05\textheight]{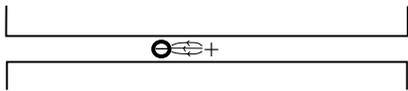}
\end{center}
\caption{A cation bound to a negative wall charge (circled). When
the cation moves away from the host the energy grows linearly with
the separation $x$. }\label{fig2}
\end{figure}

\begin{figure}[ht]
\begin{center}
\includegraphics[height=0.062\textheight]{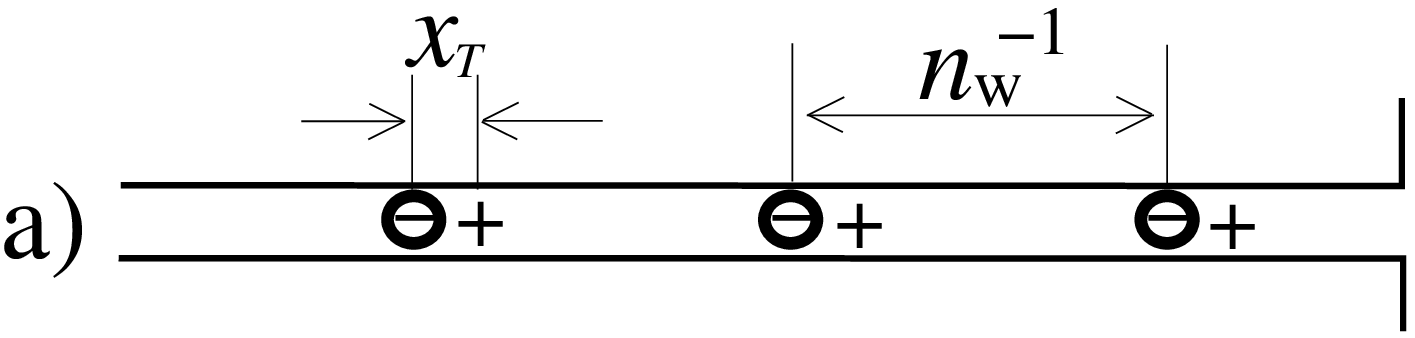} \hfill
\includegraphics[height=0.054\textheight]{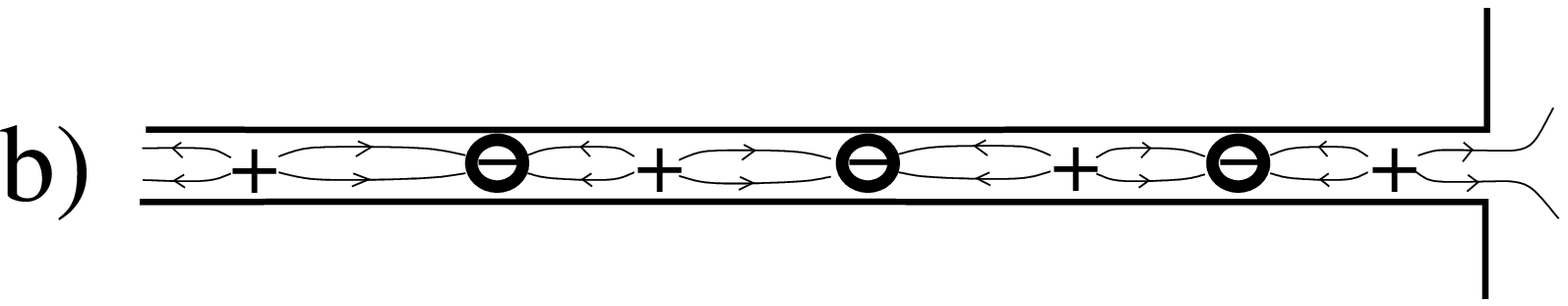}
\end{center}
\caption{The ground state and the transport saddle point of the
channel. Only right half of the channel is shown. (a) The ground
state: all cations are bound to wall charges (circled). (b) The
transport saddle point: an extra cation in the middle of the
channel (left) makes other cations free. }\label{fig3}
\end{figure}

We show below that in a large range of salt concentrations, $c$,
the wall charges attract equal number of cations from the solution
in order to make the channel neutral. Our theory is based on the
observation that the Coulomb interaction of all charges in the
short channel obeys the {\em one dimensional} Coulomb law:
$\Phi(x)\sim |x|$, same as for parallel uniformly charged planes.
Indeed, let us consider a negative charge fixed at the wall and a
cation, which arrived to the channel in order to screen it
(Fig.~\ref{fig2}). The uniform electric field between them creates
the confining ``string'' potential $\Phi(x)=eE_0|x|$. This
situation reminds two quarks confined in a meson. Condition
$\Phi(x)=k_BT$ defines the characteristic thermal length of such
classical "atom", $x_T = k_BT/eE_0 = a^{2}/l_{B}$, where $l_B
\equiv e^{2}/(\kappa_{1}k_{B}T)$ is the Bjerrum length (for water
at the room temperature $l_B=0.7\,$nm). This "atom" is similar to
an acceptor in a semiconductor (the classical length $x_T$ plays
the role of the effective acceptor Bohr radius).
It is clear that at a small dimensionless concentration of wall
charges $\gamma \equiv n_{w} x_T \ll 1$, each of them binds only
one cation (Fig.~\ref{fig3}a). Resulting neutral atoms do not
interact or overlap with each other. This system reminds a lightly
doped $p$-type semiconductor at very low temperatures when all
holes are located at their acceptors. Let us show that already for
relatively small $\gamma < 1$ the electrostatic barrier $U_L$ may
be substantially reduced.

The equilibrium partition function of the channel is $Z=(2\pi a^2
x_T/v)^{n_{w}L}$, where $v$ is the normalization cell volume. This
gives for free energy in the ground state $F_g = -
4U_{L}(0)\gamma\ln(2\pi a^2 x_T/v)$. To evaluate the transport
barrier one needs to know the free energy conditioned to the
situation where an extra cation is placed in the middle of the
channel, $F_s$. It creates the electric field $E_{0}$ which
orients the dipole moments of all "atoms" along its direction. In
other words, it orders all charges in an alternating sequence of
positive and negative ones. Due to the 1d nature of the problem,
this field unbinds each cation from its wall host and makes it
free to move between nearest neighbor wall charges
(Fig.~\ref{fig3}b). Indeed, according to the Gauss theorem the
wall charge closest to the extra cation changes electric field
from $E_0$ to $-E_0$, then its cation changes it back to $E_0$ and
so on. Thus, at any position of cations between their nearest
neighbor wall charges, the electric field is $E=\pm E_0$, i. e.
$|E|$ is constant throughout the channel. Therefore, the total
electrostatic energy is again given by Eq.~(\ref{short}). One
could think that the transport barrier is still given by
$U_{L}(0)$. This is incorrect, because the barrier is actually
determined by the difference of the {\em free} energies $F_{s} -
F_g$ of the collective transport saddle point and the ground
state. This difference is reduced by a large entropy $S$ of the
saddle point configurations.

To simplify the calculations let us imagine that the wall charges
form a periodic one-dimensional lattice along the $x$-axis. Then
$F_s= U_{L}(0)-T S$, where the entropy of the channel, enhanced by
the charge unbinding, is $S = n_{w}L\ln(\pi a^2 /2n_{w}v)$. This
gives for  $\gamma \ll 1$:
\begin{equation}
U_{L}(\gamma) = F_s - F_g  = U_L(0)\big[1 -
4\gamma\ln(1/2\gamma)\big]\, . \label{weakdopgamma}
\end{equation}

\begin{figure}[ht]
\begin{center}
\includegraphics[height=0.16\textheight]{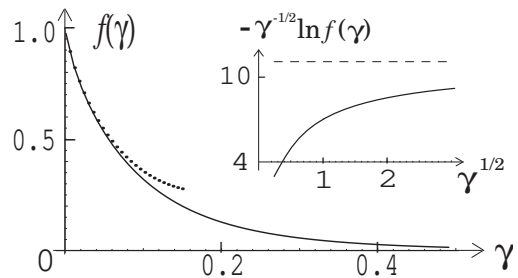}
\end{center}
\caption{The function $f(\gamma)=U_L(\gamma)/U_L(0)$. Its $\gamma
\ll 1$ asymptotics, Eq.~(\ref{weakdopgamma}), is shown by the
dotted line. The inset shows how asymptotics of Eq.~(\ref{decaya})
(dashed line) is approached at $\gamma\gg 1$.} \label{fig4}
\end{figure}

In the opposite limit, $\gamma > 1$, one may expect that atoms
overlap and destroy each other making cations free. In other
words, one could expect an insulator--to--metal (or deconfinement)
transition at some critical $\gamma \sim 1$.  This does {\em not}
happen, however. We show below that, due to the peculiar nature of
the 1d Coulomb potential, the barrier proportional to the system's
length persists to any concentration of the wall charges, no matter
how large it is. Its magnitude, though, decreases exponentially at
$\gamma \gg 1$,
\begin{equation}\label{decaya}
U_L(\gamma)/U_L(0)\propto\exp(-11.03\sqrt{\gamma})\,.
\end{equation}
Using numerical procedure outlined in the end of the paper we
calculated the ratio $f(\gamma) \equiv U_{L}(\gamma)/U_L(0)$ at
any $\gamma$ and plotted it along with the asymptotic
Eq.~(\ref{weakdopgamma}) in Fig.~\ref{fig4}.
Let us discuss the range of the salt concentration in the bulk
solution $c$, where the above results are valid. A convenient
dimensionless variable for $c$ is $\alpha \equiv \pi c a^{2}x_T$.
Eq.~(\ref{weakdopgamma}) is valid only when $\gamma \gg \alpha$,
when due to neutrality the total number of cations in the channel
is close to $n_{w}L$. In the opposite case, $\alpha > \gamma$,
additional cations together with equal number of anions enter the
channel. In order to minimize the energy all positive and negative
charges should alternate along the length of the channel or in
other words they have to be ordered. This means that each segment
of the channel between two nearest neighbor wall charges gets an
integer number $k$ of additional cation-anion pairs.

Let us now calculate the linear in  $L$ transport barrier
$U_L(\alpha,\gamma)$ at arbitrary $\alpha, \gamma \ll 1$. At the
transport saddle point, when an extra cation resides in the center
of channel and creates the field $E_0$ in both directions, all
mobile charges become free to move while keeping alternating in
charge order. The energy of all such states is again equal to
$U_L(0)$, because $|E|= E_0$ everywhere. This barrier is reduced
by a significant entropy term. To evaluate it let us first
consider entropy of a segment between two wall charges. If $k$
cation-anions pairs reside in the segment, the total number of
free ions there is $2k+1$. The total entropy of these $2k+1$ ions
is
\begin{equation}
k_B\ln [(\pi a^2 / n_{w}v)^{2k+1}/(2k+1)!] . \label{inentropyk}
\end{equation}
The factor $1/(2k+1)!$ reflects the fact that when cations and
anions are ordered it is impossible to obtain a new configuration
by exchanging them. Moving $2 k$ ions from the bulk leads to the
entropy loss $2kk_B\ln(1/cv)$ there. Subtracting this entropy from
Eq.~(\ref{inentropyk}) and taking sum over all $k$ from $0$ to
$\infty$ we obtain the total entropy of the saddle point
configuration and the renormalized barrier
\begin{equation}
U_L(\alpha, \gamma) = U_L(0)
 \left[1-4\gamma\ln\left({1\over 2\alpha}\sinh{\alpha\over
\gamma}\right)\right]. \label{correctiona}
\end{equation}
In the case $\alpha \ll \gamma$ Eq.~(\ref{correctiona}) matches
Eq.~(\ref{weakdopgamma}). In the opposite case $\alpha \gg \gamma$
Eq.~(\ref{correctiona}) crosses over to the result
$U_L(\alpha)=U_L(0)(1-4\alpha)$ obtained previously for an
uncharged  channel~\cite{Kamenev}.

So far we presented results for a periodic lattice of wall
charges. To understand the role of random distribution of wall
charges along the $x$--axis let us return to the case $\alpha \ll
\gamma \ll 1$. It is easy to show that averaging over random
nearest neighbor distances substitutes Eq.~(\ref{weakdopgamma}) by
$U_L(\gamma)=U_L(0)[1-4\gamma\ln(e^{-C}/2\gamma)]$, where
$C=0.577$ is the Euler constant. Thus, the result for randomly
distributed wall charges is similar to those for the periodic one.

Until now we concentrated on the barrier proportional to the
channel length $L$. If $\alpha \ll \gamma$ there is an additional,
independent on $L$, contribution to the transport barrier. It is
related to a large difference of concentrations of cation inside
and outside the channel. Corresponding contact (Donnan) potential
$U_D$ is created by double layers at each end consisting of one or
more negative wall charges and screening (positive) charge in
water.

For $ \gamma \ll 1$ one finds $|U_{D}| \ll U_{L}(\gamma)$ and the
channel resistance remains exponentially large. When $\gamma$
grows the barrier $U_{L}(\gamma)$ decreases and becomes smaller
than $U_{D} =- k_B T \ln(\gamma/\alpha)$, which increases with
$\gamma$ and makes the channel strongly cation selective. In this
case the measured resistance may be even smaller than the naive
geometrical diffusion resistance of the channel.

\begin{figure}[ht]
\begin{center}
\includegraphics[height=0.10\textheight]{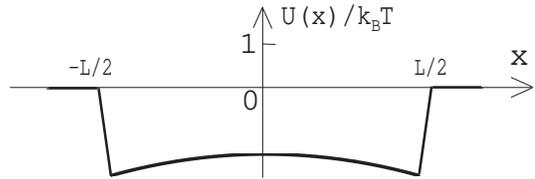}
\end{center}
\caption{The electrostatic potential for cations for the channel
with 5 wall charges considered in the text.} \label{fig5}
\end{figure}

Let us, for example, consider a channel with $L=5$ nm, $a = 0.7$
nm, $x_T=0.35$ nm at $c= 0.1$ M and $n_w = 1$ nm$^{-1}$ (5 wall
charges in the channel), which corresponds to $\alpha=0.035$ and
$\gamma = 0.35$. The bare barrier $U_L(0) = 3.5 k_{B}T$ is reduced
down to $U_L(\gamma) = 0.2 k_{B}T$. At the same time $U_{D} = -
2.5k_{B}T$. Thus due to 5 wall charges, instead of the bare
parabolic barrier of Fig.~\ref{fig1} we arrived  at the wide well
with the almost flat bottom (Fig.~\ref{fig5}).

The contact potential $U_D$ may be augmented by the negative
surface charge of the lipid membrane~\cite{Appel} or by affinity
of internal walls to a selected ion,  due to ion--specific short
range interactions~\cite{Doyle,Hille}. It seems that biological
channels have evolved to compensate large electrostatic barrier by
combined effect of $U_{D}$ and short range potentials. Our theory
is helpful if one wants to study different components of the
barrier or modify a channel. In narrow artificial nanopores there
is no reason for compensation of electrostatic barrier. In this
case, our theory may be verified by titration of wall charges.
Nanopores can also be longer than $a(\kappa_1/\kappa_2)^{1/2}$ so
that electric field lines leakage through the walls becomes
substantial. This leads to flattering of the parabolic barrier,
but its dependencies on $\alpha$ and $\gamma$ remain qualitatively
the same~\cite{Kamenev}.

Let us elaborate now on the technical aspects of the derivations.
As was first realized in Ref.~[\onlinecite{Edwards}], the
partition function of the 1d mobile Coulomb plasma may be written
as a trace of an (imaginary time) ``evolution'' operator,
$Z(q)=\mbox{Tr} \exp\{-\hat H(q)L/x_T\}$ with the Mathieu
Hamiltonian $\hat H(q)=  (i\hat \partial_\theta -q)^2 -2\alpha
\cos \theta$. The variable $q$ has a meaning of a
not-necessarily-integer screening charge induced at the channel
opening \cite{Kamenev}. The ground state of the channel
corresponds to $q=0$, while the collective saddle point, having an
uncompensated cation in the middle of the channel, to $q=1/2$. The
transport barrier is given by the difference between the two free
energies: $F_s(1/2)-F_g(0)$. As a result, the transport barrier of
an uncharged channel is proportional to the width of the lowest
Mathieu--Bloch band. It is a rapidly decreasing function  of the
mobile salt concentration $\alpha$ [\onlinecite{Kamenev}].

The immobile charges are represented by the charge ``creation''
operators $\exp\{\pm i\theta\}$, where the sign is given by the
sign of the static unit charge. For example, the partition
function of the channel with the  negative charges fixed at the
positions $x_1,x_2,\ldots$ is given by $Z(q) =\mbox{Tr}\left\{
e^{-\hat H (x_1+L/2)/x_T} e^{-i\theta} e^{-\hat H(x_2-x_1)/x_T}
e^{-i\theta}\ldots \right\}$. In the simplest case of periodically
placed charges with the dimensionless concentration $\gamma$ one
faces the spectral problem  for the {\em non--hermitian} operator
$\hat {\cal U}(q) = e^{-\hat H(q)/\gamma} e^{-i\theta}$.
The barrier may be expressed in terms of its largest eigenvalue
$e^{-\lambda^{(0)}_q}$ as
$U_L(\gamma)=k_BT(\lambda^{(0)}_{1/2}-\lambda^{(0)}_0)\gamma
L/x_T$, where $\hat {\cal U}(q)\Psi_n =
e^{-\lambda^{(n)}_q}\Psi_n$. One may demonstrate that, despite of
being non--hermitian, the operator $\hat {\cal U}(q)$ possesses
only real eigenvalues.

In the limit of  small concentration, $\gamma\ll 1$, one may write
the operator $\hat {\cal U}$ in the eigenbasis of the Hamiltonian
$\hat H$ (Bloch basis). Due to $\gamma\ll 1$ condition only lowest
eigenvalues of $\hat H$ should be retained. At $q=0$ (the band
minimum) there is a well--separated ground-state of $\hat H$,
given by $\epsilon^{(0)}_0\approx 0$. Evaluating the matrix
element of $e^{-i\theta}$, one obtains
$\lambda^{(0)}_0=-\ln(2\alpha)$.    On the other hand, at $q=1/2$
(the band maximum) and $\alpha\ll 1$ there are two almost
degenerate eigenvalues at $\epsilon^{(0,1)}_{1/2}=1/4\mp \alpha$.
Therefore,
$\lambda^{(0)}_{1/2}=1/(4\gamma)-\ln(\sinh\alpha/\gamma)$. As a
result, one arrives at Eq.~(\ref{correctiona}), which in the limit
$\alpha\ll \gamma$ yields Eq.~(\ref{weakdopgamma}).

In the large concentration limit, $\gamma> 1$, one may employ a
variant of the WKB approximation to find a spectrum of $\hat {\cal
U}$. To this end one writes the eigenfunction as $\Psi(\theta) =
e^{i\sqrt{\gamma}S(\theta)}$ and retains only the leading order in
$\gamma$. The corresponding equation for $S(\theta)$ reads as:
\begin{equation}\label{WKB}
  [S'(\theta )]^2
   + i\theta - \lambda = {\alpha\over\gamma}\left(e^{i\theta} + e^{-i\theta}
  \right) \, .
\end{equation}
The two terms on its right hand side represent positive and
negative mobile ions correspondingly. For large concentration of
fixed negative charges, $\gamma>1$, and moderate concentration of
mobile salt, $\alpha\lesssim \gamma$, one may consider only
positive ions entering the channel and disregard the negative
ones. This amounts  omitting the $e^{-i\theta}$ term on the r.h.s.
of Eq.~(\ref{WKB}) (the fixed negative charge is encoded in the
$i\theta$ term on the l.h.s.). Then shifting the $\theta$ variable
in the complex plane: $\theta\to \theta+i\ln(\gamma/\alpha)$ and
defining $\tilde \lambda \equiv \lambda - \ln (\gamma/\alpha)$,
one brings Eq.~(\ref{WKB}) into the parameterless form: $S'(\theta
)=\sqrt{\tilde\lambda-i\theta+e^{i\theta}}$. The right hand side
of this expression is an analog of the canonical momenta in the
hermitian WKB scheme. In particular its zeros play the role of the
classical turning points and determine the structure of the branch
cuts on the complex $\theta$--plane. The integrals of the
canonical momenta along such brunch cuts determine both the
spectrum and the band--width upon changing the boundary parameter
$q$.

In the strip of the complex $\theta$--plane bounded by $|\Re
\theta|<\pi$ there are two turning points at $\theta= \pm \delta
-i\delta^2/6$, where $\delta\ll 1$ is defined as $\tilde \lambda
=-1+\delta^2/2$. The structure is then periodically replicated
outside this strip. Choosing the brunch cut to run between the two
turning points and applying the Bohr--Sommerfeld quantization
rule, one finds the spectrum: $\tilde \lambda^{(n)} = -1 +
(n+1/2)\sqrt{2/\gamma}$. Since $\tilde \lambda\to -1$ in the limit
$\gamma> 1$, the two turning points are essentially close to the
origin. To find the band--width one  needs the ``tunnelling''
probability between the adjacent strips of the $\theta$--plane. To
this end one chooses the branch cuts, which emanates from the
turning points outwards, approaching $\theta=\pm \pi - i\infty$.
The (exponentiated) integral of the canonical momenta along such a
cut gives the WKB tunnelling probability and hence the band--width
$\lambda^{(0)}_{1/2}-\lambda^{(0)}_0$. Such an integration leads to
Eq.~(\ref{decaya}).

For numerical calculation it is convenient to choose the basis of
the angular momentum, $e^{im \theta}$, to evaluate $Z(q)$. In this
basis the charge creation operator $e^{-i \theta}$ takes the
matrix form $\left[ \delta_{m,m'+1} \right]$, while the
Hamiltonian $\hat H_{m,m'}= \left[ (m+q)^2 \delta_{m,m'}-\alpha
\delta_{m,m'+1}-\alpha \delta_{m,m'-1} \right]$. Truncating these
infinite matrices with  some large cutoff, one may directly
exponentiate, multiply and trace them to find the free energy for
any arrangement of fixed charges.

As emphasized by the title of this paper we study here only a very
simple model of a channel with charged walls. This is the price
for asymptotically exact
Eqs.~(\ref{weakdopgamma}),(\ref{decaya}),(\ref{correctiona}),
which show that already relatively small concentration of wall
charges dramatically reduces the self--energy transport barrier.
This conclusion is in qualitative agreement with general statement
on the role of transitional binding inside the channel on transit
probability~\cite{Bezrukov}. Our results, of course, can not
replace powerful numerical methods used for description of
specific biological channels~\cite{Roux}.

We are grateful to S. Bezrukov, A. I. Larkin and A. Parsegian for
interesting discussions. A.~K. is supported by the A.P. Sloan
foundation and the NSF grant DMR--0405212. B.~I.~S is supported by
NSF grant DMI-0210844.


\end{document}